\documentclass[twocolumn]{aastex631}

\usepackage{graphicx}
\usepackage{natbib}
\usepackage{longtable}
\usepackage{tabularx}
\usepackage{amsmath}

\newcommand{\ppxf}{\textsc{pPXF}}

\submitjournal{ApJ}

\shorttitle{SFR and dust attenuation study using CALIFA, \textit{GALEX}, and \textit{WISE}}
\shortauthors{Lee et al. 2025}
\begin{document}

\title{Spatially Resolved Star Formation Rate and Dust Attenuation of Nearby Galaxies from CALIFA, \textit{GALEX}, and \textit{WISE} Data} 

\correspondingauthor{Jong Chul Lee}
\email{jclee@kasi.re.kr}

\author[0000-0003-0283-8352]{Jong Chul Lee}
\affil{Korea Astronomy and Space Science Institute (KASI), 776 Daedeokdae-ro, Yuseong-gu, Daejeon 34055, Republic of Korea} 

\author[0000-0003-3451-0925]{Joon Hyeop Lee}
\affil{Korea Astronomy and Space Science Institute (KASI), 776 Daedeokdae-ro, Yuseong-gu, Daejeon 34055, Republic of Korea}

\author[0000-0002-0145-9556]{Hyunjin Jeong}
\affil{Korea Astronomy and Space Science Institute (KASI), 776 Daedeokdae-ro, Yuseong-gu, Daejeon 34055, Republic of Korea}

\author[0000-0002-5896-0034]{Mina Pak}
\affil{Korea Astronomy and Space Science Institute (KASI), 776 Daedeokdae-ro, Yuseong-gu, Daejeon 34055, Republic of Korea}
\affil{School of Mathematical and Physical sciences, Macquarie University, Sydney, NSW 2109, Australia}
\affil{ARC Centre of Excellence for All Sky Astrophysics in 3 Dimensions (ASTRO 3D), Australia}

\author[0000-0002-4731-9604]{Sree Oh}
\affil{Department of Astronomy and Yonsei University Observatory, Yonsei University, Seoul 03722, Republic of Korea}

\begin{abstract}
We study star formation rate (SFR) indicators and dust attenuation of 74 nearby star-forming galaxies on kiloparsec scales, based on \textit{GALEX} far-ultraviolet (FUV) and \textit{WISE} mid-infrared (MIR) images with CALIFA optical integral field spectroscopic data.
We obtain hybrid SFR indicators by combining the observed FUV and MIR luminosities and calibrate them using the dust-corrected H$\alpha$ luminosity as a reference SFR.
The simple linear combination appears to follow well the reference SFR, but the calibration residual shows a significant dependence on the specific SFR (sSFR), which can be removed by employing the combination coefficient or conversion offset that varies with the sSFR.
In the plane of gas versus stellar attenuation, the median trend line's slope ($\approx$ stellar-to-gas attenuation ratio) changes from 0.44 to 1.0 with increasing attenuation.
The differential attenuation, defined as the deviation of stellar attenuation from the median trend line, is strongly correlated with the SFR surface density and sSFR, compatible with the two-component dust model.
The differential attenuation seems to be affected by both local and global factors.
\end{abstract}


\section{INTRODUCTION}

The star formation process is closely connected with the formation and evolution of galaxies through various astrophysical phenomena such as gravitational collapse, gas inflow and outflow, and feedback mechanisms.
From the observational perspective, estimating the star formation rate (SFR) of galaxies is essential to understand this process and, by extension, the galaxy formation and evolution.
There are three types of SFR indicators that have been widely used (see \citealt{ken12} for a review).
First, ultraviolet (UV) continuum light is directly from the photosphere of young massive stars.
Second, hydrogen recombination lines (particularly H$\alpha$) are prominent in star-forming regions where young massive stars ionize the surrounding gas.
Third, infrared (IR) radiation is dominated by the thermal energy of dust grains heated by young massive stars.
Unfortunately, SFRs based on UV and optical data are significantly underestimated when galaxies suffer from severe dust attenuation, emphasizing that appropriate dust corrections are necessary to obtain accurate values.
In contrast, IR-based SFRs can be highly uncertain in dust-poor galaxies.
Moreover, IR emission depends on the physical conditions of the interstellar medium (ISM) in complex ways, so it is important to examine the reliability of SFR calibration in each IR band \citep[e.g.,][]{wu05,yua11,jar13}.

The observed H$\alpha$-to-H$\beta$ line ratio (so-called Balmer decrement) is commonly employed to correct the dust attenuation in nebular emission lines (gas attenuation).
The UV continuum slope is a good proxy for the dust attenuation of starlight (stellar attenuation), but is coupled with the stellar population age due to UV emission arising from not only young stars but also intermediate-age stars \citep[e.g.,][]{mao12,vaz16}.
The hybrid indicators, by combining UV/optical and IR data, provide robust SFR estimates as they account for both dust-obscured and unobscured star formation, inherently incorporating the dust attenuation correction.
This energy balance technique, based on the principle that energy absorbed by dust in the UV/optical range is re-emitted in the IR, has been applied to a broad range of star-forming galaxies \citep[e.g.,][]{zhu08,ken09,hao11,lee13}. 
Although spectral energy distribution fitting is a sophisticated strategy to secure intrinsic properties including dust-corrected SFR, the lack of UV or IR photometry can lead to serious biases \citep[e.g.,][]{lan13,bua14}.

Interestingly, \citet{cal94} found that stellar attenuation is a factor of two smaller than gas attenuation.
The two-component dust model proposed by \citet{cha00} has been widely accepted to explain this discrepancy.
All stars are born within dense molecular clouds (birth clouds), and some long-lived stars (with a lifetime $>$ 10 Myr) migrate to the ambient diffuse ISM.
In other words, young stars (and line-emitting gas) are deeply embedded in dusty clouds, while older stars are mixed with less dusty ISM.
As a result, the overall stellar population experiences a modest attenuation, compared to the emission lines.
The stellar-to-gas attenuation ratio ($= E(B-V)_{\rm star}/E(B-V)_{\rm gas}$) shows a considerable variation depending on the samples and methods used, typically ranging from 0.44 to 1.0 (see \citealt{pug16} and \citealt{shi20} for a summary). 

Numerous studies of SFR and dust attenuation have been based on the integrated light of individual galaxies.
However, considering the fundamental relationship between star formation and gas surface densities \citep{ken98} and the geometric effect on dust attenuation caused by the different spatial distributions of stars, gas, and dust, it is crucial to investigate local parameters in subgalactic regions.
Previous studies to calibrate spatially resolved hybrid SFR indicators are relatively abundant but are limited to a dozen galaxies \citep[e.g.,][]{cal05,boq16,euf17,bel23}.
On the other hand, detailed comparisons between spatially resolved stellar and gas attenuation are still lacking and rely on optical spectral fitting \citep{gre20,lin20,li21}.   
Although there have been attempts to break the degeneracy among stellar attenuation, age, and metallicity based on optical spectra \citep[e.g.,][]{wil17,li20}, it seems to remain challenging to accurately derive stellar attenuation from optical spectra alone (e.g., \citealt{kre13,yua18}; see also \citealt{qin22}).

With the advent of optical integral field spectroscopy (IFS) surveys like Calar Alto Legacy Integral Field Area \citep[CALIFA;][]{san12}, Sydney Australian Astronomical Observatory Multi-object Integral Field Spectrograph \citep[SAMI;][]{cro12}, and Mapping Nearby Galaxies at Apache Point Observatory \citep[MaNGA;][]{bun15}, H$\alpha$ and H$\beta$ maps become available for a large number of galaxies.
Their UV and mid-IR (MIR) counterparts can be obtained from \textit{Galaxy Evolution Explorer} \citep[\textit{GALEX};][]{mar05} and \textit{Wide-field Infrared Survey Explorer} \citep[\textit{WISE};][]{wri10} satellites, respectively.
In this work, we aim to study spatially resolved SFR and dust attenuation for as many galaxies as possible using the CALIFA, \textit{GALEX}, and \textit{WISE} datasets.
Note that CALIFA is the largest public IFS survey for galaxies at $z \lesssim$ 0.01, enabling spatially resolved studies on kpc scales together with \textit{GALEX} and \textit{WISE} data.
We derive gas attenuation from the Balmer decrement and stellar attenuation through the SFR calibration process for hybrid indicators. 
We then inspect how differently gas and stellar attenuation are influenced by several physical parameters.
The structure of this paper is as follows. 
Section \ref{sec-datamethod} describes the observational data, sample selection, and methodology used.
In Section \ref{sec-result}, we present the SFR calibration recipes and the relation between gas and stellar attenuation with related discussions.
Our findings are summarized in Section \ref{sec-sum}. 
We adopt a flat $\Lambda$CDM cosmology with
$H_0$ = 70 km s$^{-1}$ Mpc$^{-1}$, $\Omega_{\Lambda}$ = 0.7, and $\Omega_m$ = 0.3.

\section{DATA AND METHODOLOGY}\label{sec-datamethod}

\subsection{Data and Sample}\label{sec-sample}

The optical IFS data of galaxies are taken from the extended data release of CALIFA survey \citep[eCALIFA;][]{san23}.
The eCALIFA consists of 895 galaxies observed at the Calar Alto 3.5 m telescope with an exposure time of $\sim$900 s per pointing (three-pointing dithering scheme) using the V500 grating.
It covers a wavelength range of 3750--7500 \AA{} with a spectral resolution of R $\sim$850 (FWHM $\sim$6 \AA). 
The integral field unit bundle has 331 fibers with a diameter of 2.7 arcsec and its hexagonal field-of-view (FoV) is 74$\times$64 arcsec$^{2}$. 
We use the datacubes reduced with the previous pipeline (v2.2) rather than with the new pipeline (v2.3).
The major difference between the two versions is that the v2.2 datacubes have an effective spatial resolution FWHM $\sim$2$\arcsec$.5, while the v2.3 ones provide a seeing-limited resolution FWHM $\sim$1$\arcsec$.0 through an additional process of deconvolution.
The low resolution is enough in this study by considering that the datacubes are spatially matched to \textit{GALEX} and \textit{WISE} maps with FWHM = 7$\arcsec$.5 (see Section \ref{sec-datacube}). 
The galaxy parameters are adopted from the catalog of \citet{san24}: RA/Dec position, redshift, morphology, effective radius ($R_{\rm e}$), minor-to-major axis ratio ($b/a$), position angle, stellar mass, and SFR.
The galaxy mass and SFR are estimated by continuum fitting and dust-corrected H$\alpha$ luminosity, respectively, using the spectra integrated within $R_{\rm e}$.
Because these values correspond to a \citet{sal55} initial mass function (IMF), for consistency with other physical parameters in this study, we convert them to a \citet{kro01} IMF by subtracting 0.18 dex \citep{mad14}.
Considering also that the galaxy mass and SFR within $R_{\rm e}$ are about 0.3 dex smaller than the total quantities, they may appear to differ from the Salpeter IMF-based total mass and SFR in other studies by $\sim$0.5 dex.

The UV and IR data are downloaded from the z0MGS ($z$ = 0 multiwavelength galaxy synthesis) project\footnote{\href{https://doi.org/10.26131/IRSA6}
{https://doi.org/10.26131/IRSA6}} \citep{ler19,z0MGS}, which reprocessed the \textit{GALEX} and \textit{WISE} images of $\sim$16,000 nearby galaxies.
The atlas images have a common astrometry, spatial resolution, and procedure for background subtraction and foreground source masking.  
We use the \textit{GALEX} FUV ($\lambda_{\rm eff} \sim$1540 \AA), \textit{WISE} 3.4 \micron{} (W1), and 12 \micron{} (W3) intensity maps with a resolution of FWHM $=$ 7$\arcsec$.5 and a pixel size of 2$\arcsec$.75.
The rms value in the FITS header is regarded as the noise level of intensity map.    
The FUV data are provided after correcting for Galactic extinction by dust maps of \citet{sch98} and prescription of \citet{pee13}. 
We derive the stellar mass surface density ($\Sigma_*$) from the W1 intensity map, according to \citet{ler19}, assuming that the mass-to-light ratio at 3.4 \micron{} varies from 0.2 to 0.5 $M_{\sun}~L_{\sun}^{-1}$ and can be predicted using W3$-$W1 color.

\begin{figure}
    \centering
    \includegraphics[width=1.0\linewidth]{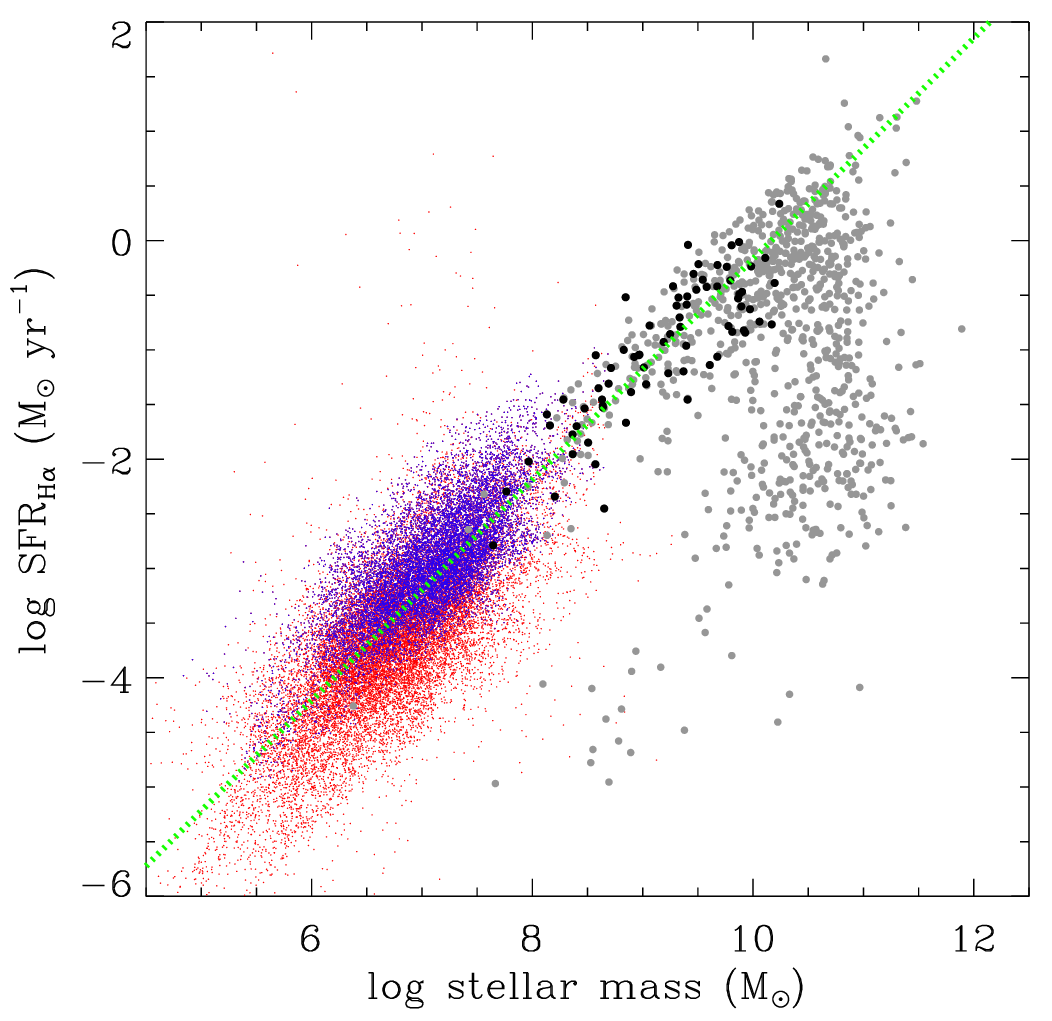} 
    \caption{Stellar mass and SFR diagram. 
    The gray and black circles represent the 895 eCALIFA galaxies and our 74 sample galaxies, respectively. 
    The red and blue dots indicate all 30,648 spaxels of the 74 galaxies and the 12,027 spaxels selected for the final analysis (see Section \ref{sec-sel}).
    The SFR is derived using dust-corrected H$\alpha$ luminosity for both galaxies and spaxels.
    The stellar mass of galaxies is from continuum fitting, while that of spaxels is based on the W1 luminosity.
    Note that the galaxy properties are obtained using the spectra integrated within $R_{\rm e}$.
    The green dotted line is the star-forming main sequence of \citet{san21}.}
    \label{fig-sample}
\end{figure}

To construct the sample for this study, we first cross-match the CALIFA galaxies with the z0MGS galaxies and obtain 613 galaxies having the FUV, optical, and MIR datasets.
Kiloparsec-scale spatial resolution has been widely used in extragalactic IFS studies as a compromise between spatial detail and sample size. 
By selecting 158 galaxies at $z <$ 0.01, we can achieve a physical resolution better than 1.53 kpc using the 7$\arcsec$.5 resolution images. 
Note that our sample galaxies are typically at $z =$ 0.0034--0.0084, corresponding to a resolution of 0.53-1.26 kpc.
We finally restrict them to 74 late-type galaxies (from Sa to Irr) with $b/a >$ 0.4 to mitigate the complexity of interpretation.
The early-type galaxies are excluded because they rarely show on-going star formation activity and the vast majority of them do not fall into the star-forming main sequence \citep[SFMS;][]{elb07}.
Moreover, their recent star formation appears to have an external origin, unlike typical late-type galaxies \citep[e.g.,][]{sal10,jeo22,lee23}. 
The galaxies with companions within 2$R_{\rm e}$ are also eliminated to avoid considering environmental effects and to get clean radial profiles (see Section \ref{sec-global}).
The highly inclined galaxies with $b/a \lesssim$ 0.4, corresponding to an inclination angle $\gtrsim 66 \arcdeg$, are optically thick and their physical properties, including SFR and dust attenuation, are difficult to accurately estimate \citep[e.g.,][]{xia12,li19,doo22}. 
Figure \ref{fig-sample} shows the distribution of 74 sample galaxies (black circles) in the plane of stellar mass and SFR.
They are star-forming galaxies, and most of them (84 percent) closely follow the SFMS (green dotted line), defined by \citet{san21}, within $\pm$0.5 dex.
There are no starburst or quiescent sample galaxies that deviate by more than a factor of ten above or below the sequence.

\subsection{Datacube Reconstruction}\label{sec-datacube}

For a direct comparison between the optical and FUV/MIR datasets, we reconstruct the CALIFA datacubes to be spatially matched with the z0MGS images.
First, the bad pixels marked in the \textsc{BADPIX} extension are replaced by the median value of 20 adjacent good pixels.
The datacubes are then smoothed with the suitable convolution kernel of \citet{ani11} to have the common spatial resolution of Gaussian FWHM = 7$\arcsec$.5.
Note that the point spread functions of the original datacubes are well described by a 2D Moffat profile with FHWM = 2.50 $\pm$ 0.34 arcsec \citep{san16}.
The pixel size of datacubes is changed from 1$\arcsec$.0 to 2$\arcsec$.75 using the \textsc{IDL/REBIN} function.
Finally, the datacubes are subpixel shifted by a linear interpolation to correct the residual misalignment.
This process is carried out for each wavelength element (per 2 \AA) and the resulting datacube has 1876 images with 28$\times$28 pixels (77$\times$77 arcsec$^{2}$).
We also generate the error-cubes via the same process, but taking into account the error propagation in spatially co-added spectra (see the appendix of \citet{gar15} for more details).

Data could be severely distorted in the spaxels near the edge of CALIFA FoV by the smoothing procedure because a significant portion of the convolution kernel is outside the FoV.
We exclude the spaxels where more than 15 percent of the kernel is outside the FoV from further analysis.
The spaxels with fewer than 12 elements having a signal-to-noise ratio (S/N) per \AA{} $>$ 5 are also rejected, considering that even the major emission lines would not be detected.
Out of the 74 datacubes, a total of 30,648 spaxels are retained for further analysis.
The 1$\sigma$ range of their median S/Ns is 3.9--25.5.

\subsection{Spectral Fitting}\label{sec-ppxf}

To extract spectral information from each spaxel, we employ the penalized pixel-fitting method \citep[\ppxf;][]{cap04,cap23} and the E-MILES stellar population synthesis templates \citep{vaz16}.
A set of 78 templates is adopted, covering 13 ages (0.06--15.8 Gyr, logarithmically spaced in steps of 0.2 dex) and 6 metallicites ([M/H] = $-$1.71, $-$1.31,$-$0.71,$-$0.40, 0.00, and 0.22).
The templates have a resolution of FWHM = 2.51 \AA, so they are convolved to match the CALIFA resolution of FWHM = 6.0 \AA.
Before running \ppxf, the input spectra are transformed to the rest-frame and are corrected for Galactic extinction using \citet{car89} curve and \citet{sch98} maps.
The spectral fitting is performed in three steps following the recommendation of \citet{ems22}: the first is for the stellar kinematics, the second is for the stellar population properties, and the third is for the measurement of ionized gas emission lines.
The bad pixels are removed by iterative sigma clipping and the regions around major emission lines are masked (width = $\pm$400 km s$^{-1}$) in the first and second steps.
The stellar kinematic parameters (velocity, velocity dispersion, $h$3, and $h$4) are derived in the first step and are kept fixed in the second and third steps.
The additive Legendre polynomials of 10th order are used in the first step, while the multiplicative polynomials are adopted in the second and third steps.
The regularization option is turned off throughout the process.

\begin{figure*}
    \centering
    \includegraphics[width=1.0\linewidth]{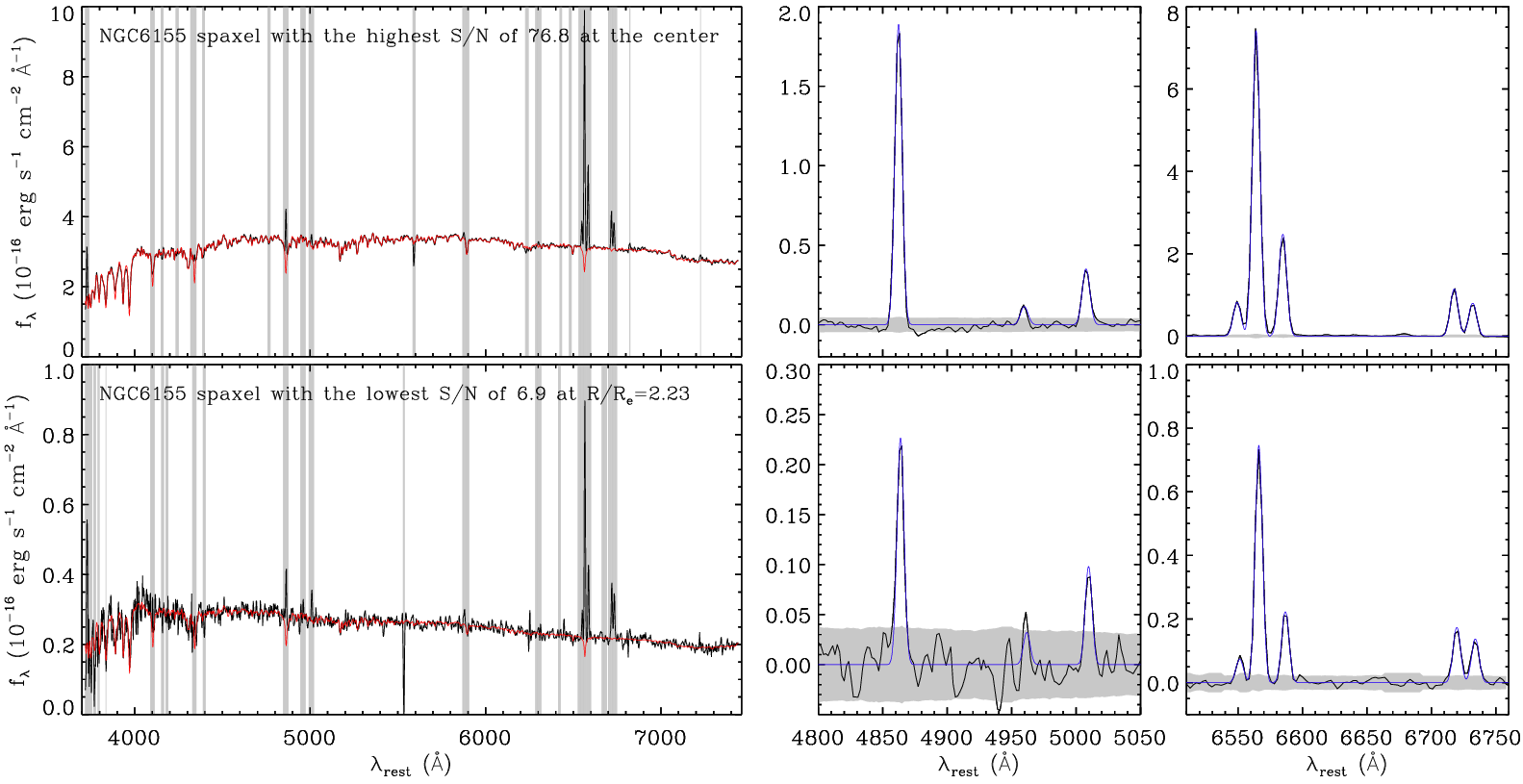}
    \caption{Examples of spectral fitting for NGC 6155 spaxels.
    The top and bottom panels display the highest and lowest S/N cases among the 315 final spaxels in NGC 6155 (see Section \ref{sec-sel}), respectively.
    In the left panels, the observed spectrum (rest-frame) and the stellar component fit are shown in black and red lines. 
    The gray shaded regions denote the bad pixels and line masks.
    The middle and right panels present the zoomed-in views of the H$\beta$ and H$\alpha$ regions, respectively. 
    The black and blue lines indicate the continuum-subtracted spectrum and the emission line fits.
    The gray shaded area stands for the error level.}
    \label{fig-spec}
\end{figure*}

\begin{figure*}
    \centering
    \includegraphics[width=1.0\linewidth]{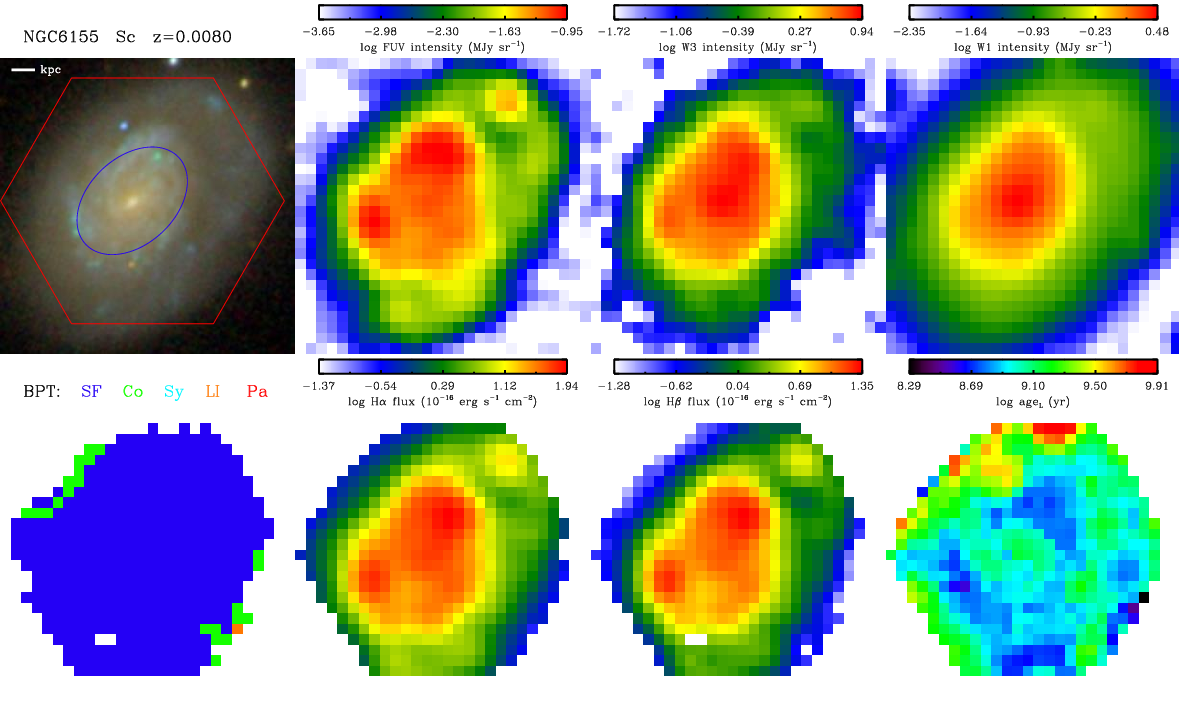} 
    \caption{Example maps of NGC 6155. The $gri$ composite image (from the Sloan Digital Sky Survey; \citealt{alm23}) is shown in the top-leftmost panel with
    the morphology and redshift information. The red hexagon, blue ellipse, and white bar are superposed 
    to present the CALIFA FoV, effective aperture, and physical scale, respectively.
    The other top panels are the z0MGS FUV, W3, and W1 intensity maps.
    The bottom panels are our outputs from CALIFA datacube reconstruction and spectral fitting. 
    From left to right, BPT classification (star-forming: blue, composite: green, Seyfert: cyan, LINER: orange, and passive: red), H$\alpha$ flux, H$\beta$ flux, and age$_{\rm L}$ maps are displayed.
    All maps have a size of 77$\times$77 arcsec$^{2}$. North is up and east is to the left.}
    \label{fig-maps}
\end{figure*}

In the third step, the stellar continuum and emission lines are fitted simultaneously via the addition of Gaussian templates.
The kinematics of emission lines considered are tied together into three groups: 
  hydrogen Balmer lines (H$\delta$, H$\gamma$, H$\beta$, and H$\alpha$), 
  low ionization lines ([O~{\sc ii}]$\lambda\lambda$3726,29, [O~{\sc i}]$\lambda\lambda$6300,64, 
  [N~{\sc ii}]$\lambda\lambda$6548,84, and [S~{\sc ii}]$\lambda\lambda$6717,31), and
  high ionization lines ([O~{\sc iii}]$\lambda\lambda$4959,5007).
The doublet line ratios of [O~{\sc iii}]$\lambda$4959/[O~{\sc iii}]$\lambda$5007, [O~{\sc i}]$\lambda$6364/[O~{\sc i}]$\lambda$6300, and [N~{\sc ii}]$\lambda$6548/[N~{\sc ii}]$\lambda$6584 are constrained to be 1/3.  
Among the various output parameters, the H$\beta$, [O~{\sc iii}]$\lambda$5007, H$\alpha$, and [N~{\sc ii}]$\lambda$6584 fluxes (and their errors) and the luminosity-weighted age of stellar populations (age$_{\rm L}$) are used in this study.
Figures \ref{fig-spec} and \ref{fig-maps} display the spectral fitting results of NGC 6155 spaxels and the H$\alpha$ and H$\beta$ flux maps of NGC 6155 along with other information, as an example.

\subsection{Spaxel Selection}\label{sec-sel}

Among the 30,648 spaxels, we choose 15,966 spaxels with S/N $>$ 10 for H$\alpha$ and H$\beta$ fluxes, S/N $>$ 5 for FUV intensity, and W3 intensity $>$ 0.
Because the line flux errors are taken from the \ppxf{} output without bootstrapping simulations, they can be much smaller than the actual values.
Therefore, we empirically determine the S/N threshold of 10 for the H$\alpha$ and H$\beta$ fluxes to ensure line measurements.
We do not impose any S/N cut on the W3 intensity to prevent a significant decrease in the number of available spaxels\footnote{The data used in this study appear to be stronger in the order of H$\alpha$, FUV, H$\beta$, and W3, based on their S/N distributions. Applying an S/N cut of 2 (5) to the W3 intensity reduces the final sample size by 5 (18) percent.} and bias in the relation between gas and stellar attenuation (see Section \ref{sec-ebv}).
Note that our main results remain unchanged even if an S/N $>$ 5 criterion for the W3 intensity is added.

We classify the ionizing source of spaxels in the standard Baldwin-Phillips-Terlevich diagram \citep[BPT;][]{bal81} of [O~{\sc iii}]$\lambda5007$/H$\beta$ versus [N~{\sc ii}]$\lambda6584$/H$\alpha$.
Star-forming spaxels lie below the limit of pure star formation of \citet{kau03}.
Composite spaxels (both from star formation and active galactic nuclei) are between the pure star formation line and the (theoretical) maximum starburst line of \citet{kew01}.
The spaxels above the maximum starburst line are divided into Seyfert and LINER (low-ionization nuclear emission line region) spaxels based on the criterion of \citet{cid10}.
The spaxels with H$\alpha$ equivalent width $<$ 1 \AA{} are considered passive, regardless of their position on the diagram.
To minimize contamination by regions not associated with young stars, 15,511 star-forming spaxels are selected.
These star-forming spaxels can contain not only H~{\sc ii} regions but also diffuse ionized gas (DIG) regions, which are powered by radiation leaking from H~{\sc ii} regions \citep[e.g.,][]{haf09,bel22}.  
According to the guideline of \citet{zha17}, H$\alpha$ surface density $\Sigma_{\rm H\alpha}$ $<$ 10$^{39}$ erg s$^{-1}$ kpc$^{-2}$, about 35 percent of our spaxels are DIG-dominated.
However, since a bimodal distribution of $\Sigma_{\rm H\alpha}$ is not found in this study, H~{\sc ii} and DIG regions are not considered separately.

We finally use 12,027 spaxels with H$\alpha$/H$\beta$ ratio $>$ 2.86 to focus on dusty star-forming regions.
The Balmer decrement is generally larger than its intrinsic value of 2.86 (for case B recombination with $T =$ 10,000 K and $n_e =$ 100 cm$^{-3}$; \citealt{ost06}) due to the dust reddening effect.  
The intrinsic value is known to change little with electron density but to depend weakly on temperature (from 3.04 to 2.75 for 5000--20,000 K; \citealt{dop03}).
It would be interesting to inspect how much this value varies in different parts of galaxies and why a substantial fraction of spaxels has H$\alpha$/H$\beta$ $<$ 2.86.
However, it is beyond the scope of this study and the fixed value of 2.86 is adopted.
The stellar mass ($M_*$ = $\Sigma_*~\times$ physical area of each spaxel) and SFR (see Section \ref{sec-sfr}) of spaxels are overplotted in Figure \ref{fig-sample}.
The 12,027 selected spaxels (blue dots) are found along the SFMS, like the sample galaxies.
Note that the median S/Ns of these spaxels are typically 7.9–35.8.

\section{RESULTS AND DISCUSSION}\label{sec-result}

\subsection{Star Formation Rate Indicators}\label{sec-sfr}

We regard the H$\alpha$ luminosity of each spaxel as a reference SFR indicator.
To derive the intrinsic H$\alpha$ luminosity, we first estimate the color excess toward gas components as 
  \begin{equation}
  E(B-V)_{\rm gas} = \frac{2.5}{k_{\rm H\beta}-k_{\rm H\alpha}}~\log \left[\frac{({\rm H\alpha}/{\rm H\beta})_{\rm obs}}{2.86}\right],
  \end{equation}
  where $k_{\rm H\alpha}$ and $k_{\rm H\beta}$ are the reddening curve values of \citet{cal00} with $R_{\rm V}$ = 4.05 at the H$\alpha$ and H$\beta$ wavelengths. 
The reddening curve of \citet{car89} has also been used in previous studies.
Although the difference between the commonly used curves is not significant in the optical, the \citet{cal00} curve seems to be more suitable for IR-selected systems \citep[e.g.,][]{lee13,qin19}.
We then correct for the dust effect as
  \begin{equation}
  L_{\rm H\alpha,corr} = L_{\rm H\alpha}~10^{~0.4~k_{\rm H\alpha}~E(B-V)_{\rm gas}}.
  \end{equation}

The FUV and W3 luminosities are computed at the effective wavelengths (monochromatic; $\nu L_{\nu}$).
The stellar continuum peaks around the near-IR but remains even in the MIR.
To obtain the MIR dust luminosity for a reliable SFR indicator, the MIR emission contributed by stellar populations should be removed. 
Using the spaxels with age$_{\rm L} >$ 10 Gyr in the CALIFA early-type galaxies, we find that their W3-to-W1 intensity ratio is on average 0.18, which is compatible with the scale factor of $\sim$0.15 in \citet{jar13}.
With the assumption that the W1 emission is dust-free and the pure starlight has a W3/W1 ratio of 0.18, the W3 dust luminosity is derived by subtracting a scaled W1 luminosity from the entire W3 luminosity.
It decreases the W3 luminosity of dusty star-forming spaxels by $\sim$9 percent but has little effect on our conclusions.
We hereafter denote $L_{\rm W3}$ as the W3 dust luminosity. 

\begin{figure*}
    \centering
    \includegraphics[width=1.0\linewidth]{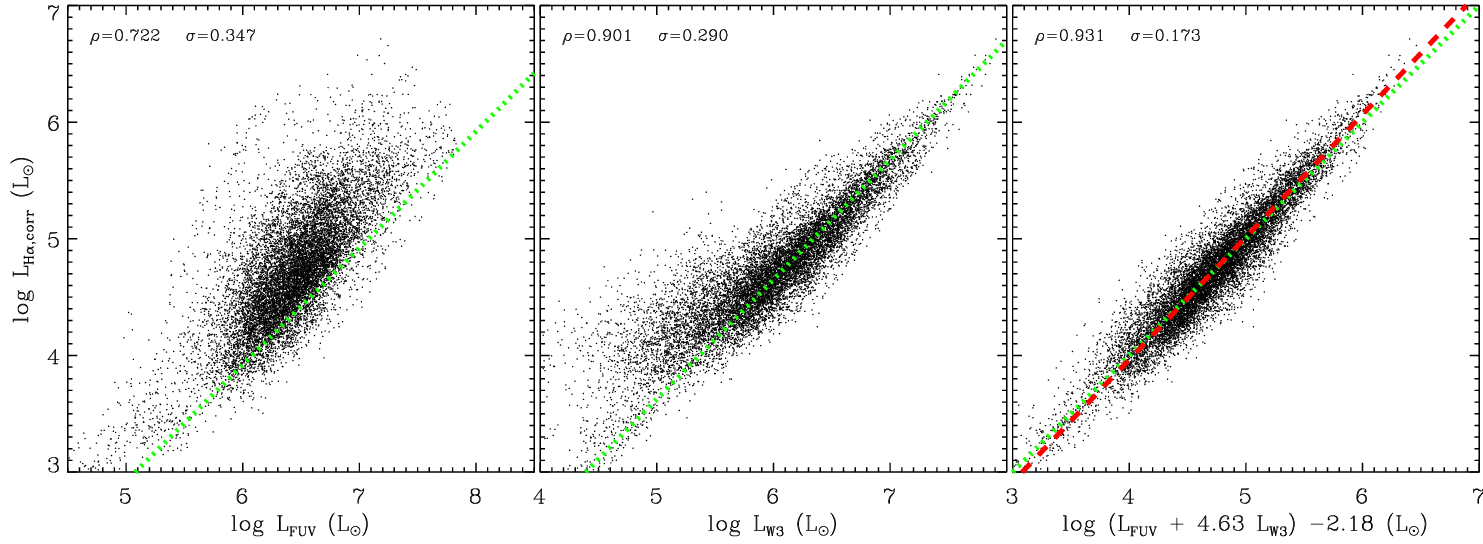} \\
    \caption{Dust-corrected H$\alpha$ luminosity versus FUV (left), 12 \micron{} (middle), and their combined (right) luminosities. 
    The Spearman's $\rho$ and standard deviation $\sigma$ between the two parameters are presented in the top-left corner of each panel.
    The green dotted lines in the left and middle panels are based on the SFR conversion factors in \citet{ken12} and the star-forming galaxy relation of \citet{lee13}, respectively.
    In the right panel, the green dotted line means the one-to-one relation, and the red dashed line is from the bisector fit to the data.}
    \label{fig-sfr-comb}
\end{figure*}

In the left panel of Figure \ref{fig-sfr-comb}, we compare $L_{\rm H\alpha,corr}$ and $L_{\rm FUV}$.
Although both parameters are SFR indicators, there is a relatively moderate correlation (Spearman rank correlation coefficient $\rho$ = 0.722) and a large scatter (standard deviation $\sigma$ = 0.347 dex) between the two.
This originates from the lack of dust correction for $L_{\rm FUV}$ in the sense that most of our spaxels are located to the left of the relation from the two conversion factors (H$\alpha$ and FUV luminosities to SFR) in \citet{ken12}, particularly in the more luminous (i.e., higher SFR) regime. 
In the middle panel, the spaxels closely follow the line extrapolated from the star-forming galaxy fit of \citet{lee13}, with $\rho$ = 0.901 and $\sigma$ = 0.290, indicating that $L_{\rm W3}$ is a better SFR indicator than (dust-uncorrected) $L_{\rm FUV}$.
However, the low SFR spaxels appear to have relatively small values of $L_{\rm W3}$/$L_{\rm H\alpha,corr}$.
A similar phenomenon has been reported in previous studies \citep[e.g.,][]{lee13,clu14,bro17}.
It is mainly because metal-poor regions have a low dust-to-gas ratio \citep[e.g.,][]{rem14,dev19} or a low abundance of polycyclic aromatic hydrocarbon (PAH) molecules \citep[e.g.,][]{dra07,gal08}, resulting in a deficit of $L_{\rm W3}$ containing a prominent PAH feature at 11.3 \micron.

By combining $L_{\rm FUV}$ and $L_{\rm W3}$, we can obtain a tight relation over a wide range with $\rho$ = 0.931 and $\sigma$ = 0.173, as shown in the right panel.
The linear combination coefficient of 4.63 is determined to minimize the scatter of the relation.
This value of 4.63 seems reasonable considering the combination coefficients for FUV and MIR in previous studies: 4.27 for W3 \citep{ler19}, 6.1 for \textit{Spitzer} 24 $\micron$ \citep{ler08}, 3.89 for \textit{IRAS} 25 $\micron$ \citep{hao11}, and 4.52 for \textit{WISE} 22 $\micron$ \citep{cat15}. 
The offset of $-$2.18 dex between the two SFR indicators is from the median of log $L_{\rm H\alpha,corr}$/($L_{\rm FUV}$ + 4.63 $L_{\rm W3}$).
Our value of $-$2.18 is comparable to the offsets between SFR conversion factors for H$\alpha$ and FUV luminosities in \citet[][from $-$2.04 to $-$2.15]{hao11}, \citet[][$-$2.08]{ken12}, and \citet[][$-$2.16]{bel23}.
Although this simple linear combination produces a good one-to-one relationship, the resultant relation is found to be slightly tilted (slope = 1.051 $\pm$ 0.003) using the bisector method of \citet{iso90}.
It implies that the combination coefficient or the conversion offset depends on SFR.

\begin{figure*}
    \centering
    \includegraphics[width=0.8\linewidth]{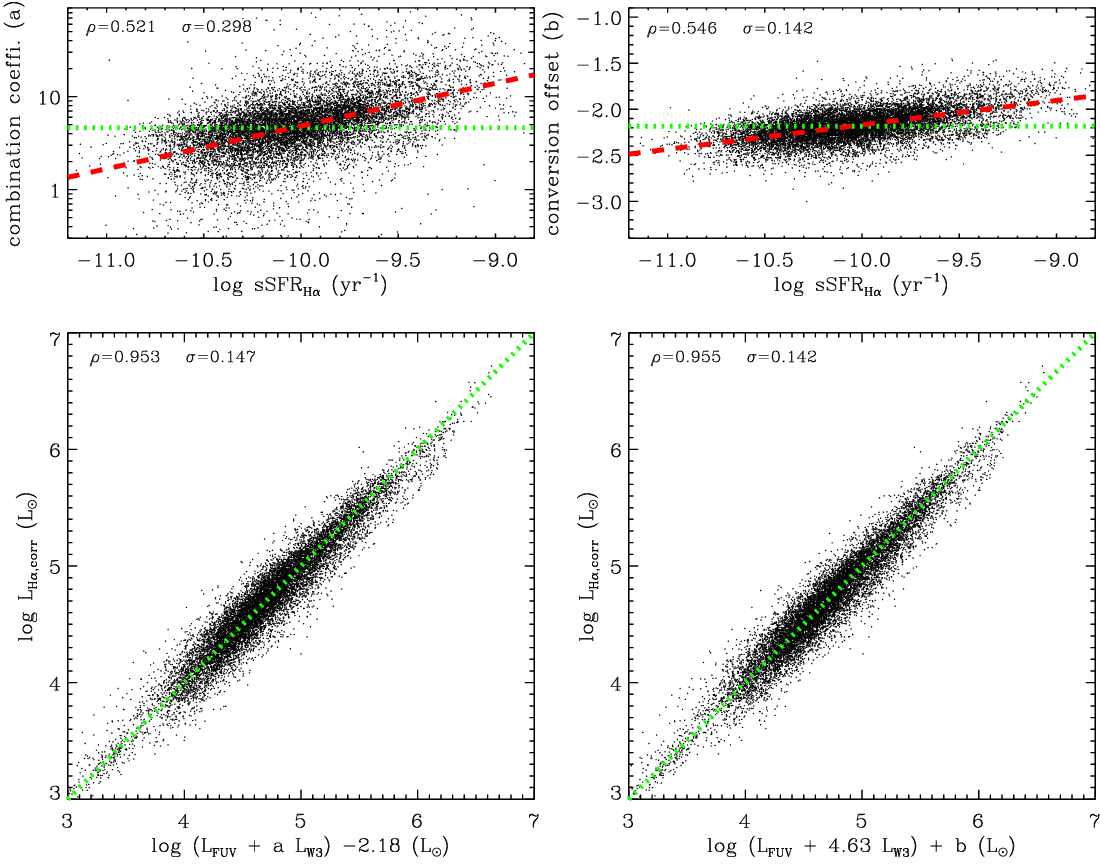} \\
    \caption{Top: coefficient for FUV and W3 combination (left) and offset between SFR conversion factors for H$\alpha$ and FUV (right) as a function of sSFR$_{\rm H\alpha}$. 
    The green dotted and red dashed lines denote their median values (4.63 and $-$2.18) and the best fits to the data, respectively.
    The numbers in the top-left corner are the Spearman's $\rho$ and the fitting residual $\sigma$ (dex).
    Bottom: dust-corrected H$\alpha$ luminosity versus FUV and W3 combined luminosity using the coefficient \textit{\textbf{a}} (left) and the offset \textit{\textbf{b}} (right) predicted by sSFR$_{\rm H\alpha}$.
    The green dotted line is the one-to-one relation.}
    \label{fig-sfr-corr}
\end{figure*}

We derive the SFR of each spaxel from $L_{\rm H\alpha,corr}$ following \citet{ken12}:
  \begin{equation}
  {\rm SFR_{H\alpha}}~(M_{\odot}~{\rm yr}^{-1})=2.07 \times 10^{-8}~ L_{\rm H\alpha,corr}~(L_{\odot}).
  \end{equation}
When the conversion offset is fixed to $-$2.18, the combination coefficient can be given by (10$^{2.18}$ $L_{\rm H\alpha,corr} - L_{\rm FUV}$)/$L_{\rm W3}$ from log $L_{\rm H\alpha,corr}$ = log ($L_{\rm FUV}$ + \textit{\textbf{x}} $L_{\rm W3}$) $-$2.18.
This coefficient strongly depends on the specific SFR (sSFR$_{\rm H\alpha}$ = SFR$_{\rm H\alpha}$/$M_*$) rather than other parameters such as SFR$_{\rm H\alpha}$, age$_{\rm L}$, $\Sigma_*$, and $E(B-V)_{\rm gas}$.
In the top-left panel of Figure \ref{fig-sfr-corr}, we plot the coefficient as a function of sSFR$_{\rm H\alpha}$ and obtain its power-law relation using an ordinary least squares fit.
We hereafter denote the combination coefficient predicted by sSFR$_{\rm H\alpha}$ as \textit{\textbf{a}}:
  \begin{equation}
  {\rm log~\textit{\textbf{a}}} = 5.28 + 0.46~{\rm log~sSFR_{H\alpha}~(yr^{-1})}.
  \end{equation}
The tendency for the coefficient to increase with sSFR is probably induced by the fact that the diffuse IR emission component from dust heated by old stellar populations becomes important at low sSFR environments (e.g., \citealt{cor08,boq21}; see also \citealt{pat24}). 
\citet{bel23} assumed a broken power-law by expecting that the old star contribution can be ignored in high sSFR regions and argued that their coefficients are constant at log sSFR (yr$^{-1}$) $\gtrsim$ $-$10, but we do not find such a flattening in the relation.

On the other hand, when fixing the combination coefficient to 4.63, the conversion offset is given by log $L_{\rm H\alpha,corr}$/($L_{\rm FUV}$ + 4.63 $L_{\rm W3}$) and is also correlated with sSFR$_{\rm H\alpha}$ as shown in the top-right panel.
The conversion offset predicted by sSFR$_{\rm H\alpha}$ is referred to as \textit{\textbf{b}}:
  \begin{equation}
  \textit{\textbf{b}} = 0.48 + 0.27~{\rm log~sSFR_{H\alpha}~(yr^{-1})}.
  \end{equation}
The SFR conversion factors vary depending on assumptions regarding star formation history (SFH), metallicity, and IMF (see \citealt{ken12} for detailed discussions).
According to simulations by \citet{hao11}, the effect of SFH is more pronounced in the FUV than in the H$\alpha$.
This is due to intermediate-age stellar populations contributing to the FUV emission: the H$\alpha$ and FUV trace star formation time scales of up to $\sim$10 and $\sim$100 Myr, respectively.
The observational study of \citet{ler19} revealed that the conversion factor from FUV luminosity to SFR actually increases with sSFR, which is dependent on SFH. 

In the bottom panels of Figure \ref{fig-sfr-corr}, we present $L_{\rm H\alpha,corr}$ versus $L_{\rm FUV}$ and $L_{\rm W3}$ combinations using the coefficient \textit{\textbf{a}} or the offset \textit{\textbf{b}}.
Compared to the simple linear combination relation in the right panel of Figure \ref{fig-sfr-comb}, these relations do not tilt and are stronger and tighter.  
In reality, both the coefficient and the offset vary with sSFR, but it is difficult to estimate these variations simultaneously using our data alone.
Fixing one while deriving the other is an alternative approach, and the two results appear to be practically equivalent in this study.
However, we prefer to adopt the variable \textit{\textbf{b}} over \textit{\textbf{a}} because it makes the relation stronger and tighter, albeit slightly.

In summary, our recipes for converting $L_{\rm FUV}$ and $L_{\rm W3}$ into an accurate SFR indicator as $L_{\rm H\alpha,corr}$ are  
  \begin{eqnarray}
  L_{\rm FUV+W3,const}  &=&~L_{\rm FUV} + 4.63~L_{\rm W3},\\
  L_{\rm FUV+W3,tilt}~~ &=&~0.45~(L_{\rm FUV} + 4.63~L_{\rm W3})^{1.05},\\  
  L_{\rm FUV+W3,coeffi} &=&~L_{\rm FUV} + \textit{\textbf{a}}~L_{\rm W3},\\
  L_{\rm FUV+W3,offset} &=&~10^{~\textit{\textbf{b}}+2.18}~(L_{\rm FUV} + 4.63~L_{\rm W3}),
  \end{eqnarray}
  where log \textit{\textbf{a}} = 5.28 + 0.46 log sSFR$_{\rm H\alpha}$ (yr$^{-1}$) and \textit{\textbf{b}} = 0.48 + 0.27 log sSFR$_{\rm H\alpha}$ (yr$^{-1}$).
Note that these recipes are based on regions with $-$4.0 $\lesssim$ log SFR ($M_{\odot}$ yr$^{-1}$) $\lesssim$ $-$1.5 but are applicable to typical star-forming galaxies as well, which are useful to obtain their SFR: 
  \begin{equation}
  {\rm SFR_{FUV+W3}}~(M_{\odot}~{\rm yr}^{-1})=1.37 \times 10^{-10}~ L_{\rm FUV+W3}~(L_{\odot}).
  \end{equation}
Among these four cases, we regard $L_{\rm FUV+W3,const}$ (from the simple linear combination) and $L_{\rm FUV+W3,offset}$ (using the sSFR-dependent offset) as the most basic and advanced, respectively.

\begin{figure*}
    \centering
    \includegraphics[width=1.0\linewidth]{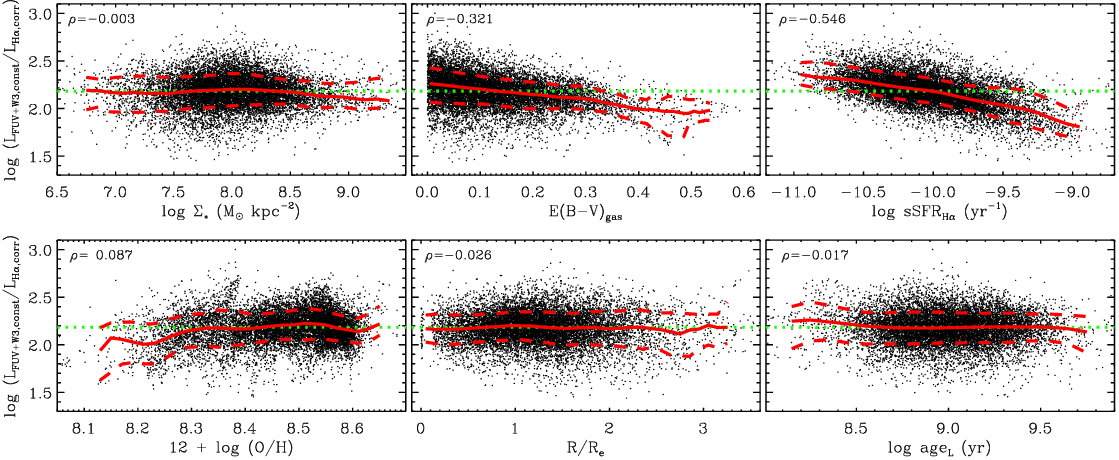} 
    \caption{Ratio between $L_{\rm FUV+W3,const}$ (simple linear combination) and $L_{\rm H\alpha,corr}$ as a function of $\Sigma_*$ (top-left), $E(B-V)_{\rm gas}$ (top-middle) sSFR$_{\rm H\alpha}$ (top-right), 12 + log (O/H) (bottom-left), $R$/$R_{\rm e}$ (bottom-middle), and age$_{\rm L}$ (bottom-right).
    In each panel, the red solid and dashed lines represent the sliding medians and 16th--84th percentile distributions, respectively.
    The green horizontal dotted line indicates the median ratio of 2.18 dex.  
    The Spearman's $\rho$ is denoted in the top-left corner.}
    \label{fig-sfr-res1}
\end{figure*}

\begin{figure*}
    \centering
    \includegraphics[width=1.0\linewidth]{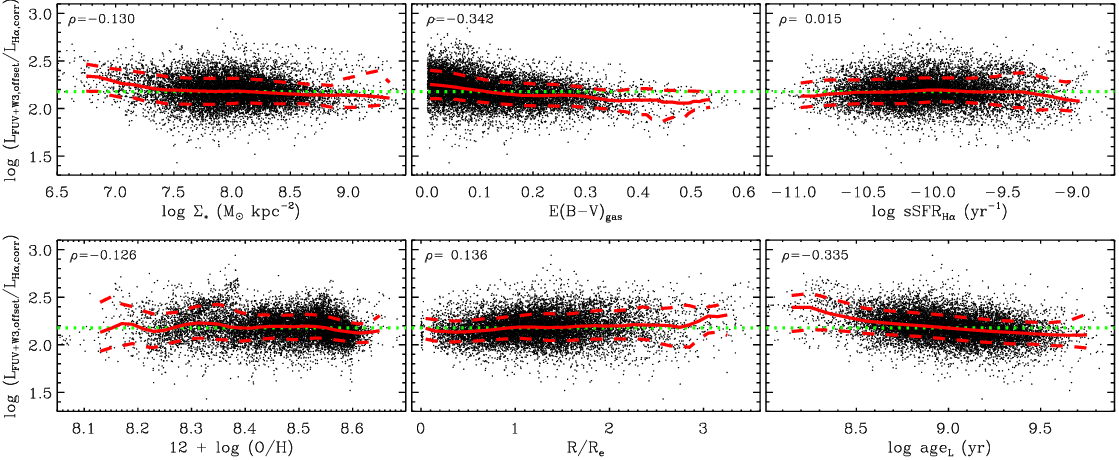} \\
    \caption{Same as Figure \ref{fig-sfr-res1}, but for $L_{\rm FUV+W3,offset}$, using the fixed coefficient and the variable offset \textit{\textbf{b}}.}
    \label{fig-sfr-res2}
\end{figure*}

To examine how well $L_{\rm FUV+W3,const}$ and $L_{\rm FUV+W3,offset}$ follow the reference SFR indicator without dependence on physical properties, we present $L_{\rm FUV+W3,const}$/$L_{\rm H\alpha,corr}$ and $L_{\rm FUV+W3,offset}$/$L_{\rm H\alpha,corr}$ as a function of $\Sigma_*$, $E(B-V)_{\rm gas}$, sSFR$_{\rm H\alpha}$, 12 + log (O/H), $R$/$R_{\rm e}$, and age$_{\rm L}$ in Figures \ref{fig-sfr-res1} and \ref{fig-sfr-res2}.
The oxygen abundance, a proxy of gas-phase metallicity, is estimated using the O3N2 index of [O~{\sc iii}]$\lambda5007$/H$\beta$ $\times$ H$\alpha$/[N~{\sc ii}]$\lambda6584$ and the calibration of \citet{mar13}.
The $R$/$R_{\rm e}$ indicates the galacto-centric distance normalized to $R_{\rm e}$.
In the $L_{\rm FUV+W3,const}$/$L_{\rm H\alpha,corr}$ plots, the notable correlations\footnote{The distribution of $\rho$ between randomly shuffled parameter sets has a standard deviation of $\sim$0.01. We consider the correlation to be notable when $\left\vert \rho \right\vert$ is greater than 0.1.} are found only with $E(B-V)_{\rm gas}$ and sSFR$_{\rm H\alpha}$. 
However, the correlation with $E(B-V)_{\rm gas}$ does not seem to be meaningful in the sense that $L_{\rm H\alpha,corr}$ is strongly affected by $E(B-V)_{\rm gas}$ and there is no correlation between $L_{\rm FUV+W3,const}$/$L_{\rm H\alpha,corr}$ and stellar attenuation $E(B-V)_{\rm star}$.
The strongest correlation with sSFR$_{\rm H\alpha}$ causes the tilt of relation between $L_{\rm H\alpha,corr}$ and $L_{\rm FUV+W3,const}$ and is removable by using the conversion offset (or the combination coefficient) that varies with sSFR$_{\rm H\alpha}$.
We confirm that $L_{\rm FUV+W3,offset}$ has changed from a recent star formation tracer like $L_{\rm FUV+W3,const}$ to a current star formation tracer like $L_{\rm H\alpha,corr}$, based on the negligible $\rho$ between $L_{\rm FUV+W3,offset}$/$L_{\rm H\alpha,corr}$ and sSFR$_{\rm H\alpha}$.
However, it generates a moderate dependence of $L_{\rm FUV+W3,offset}$/$L_{\rm H\alpha,corr}$ on age$_{\rm L}$ because sSFR$_{\rm H\alpha}$ is entangled with age$_{\rm L}$ ($\rho = -$0.448).
The weak dependence of $L_{\rm FUV+W3,offset}$/$L_{\rm H\alpha,corr}$ on $\Sigma_*$, 12 + log (O/H), and $R$/$R_{\rm e}$ can be understood by considering that sSFR$_{\rm H\alpha}$ is a normalized quantity by $M_*$ (= $\Sigma_*~\times$ spaxel area) and $\Sigma_*$ is intimately linked with 12 + log (O/H) ($\rho$ = 0.671) and $R$/$R_{\rm e}$ ($\rho = -$0.610).
In the view that $L_{\rm FUV+W3,offset}$/$L_{\rm H\alpha,corr}$ does not have a severe dependency on any physical parameters, we regard $L_{\rm FUV+W3,offset}$ as a better SFR indicator than $L_{\rm FUV+W3,const}$.

\subsection{Gas versus Stellar Attenuation}\label{sec-ebv}

\begin{figure}
    \centering
    \includegraphics[width=0.9\linewidth]{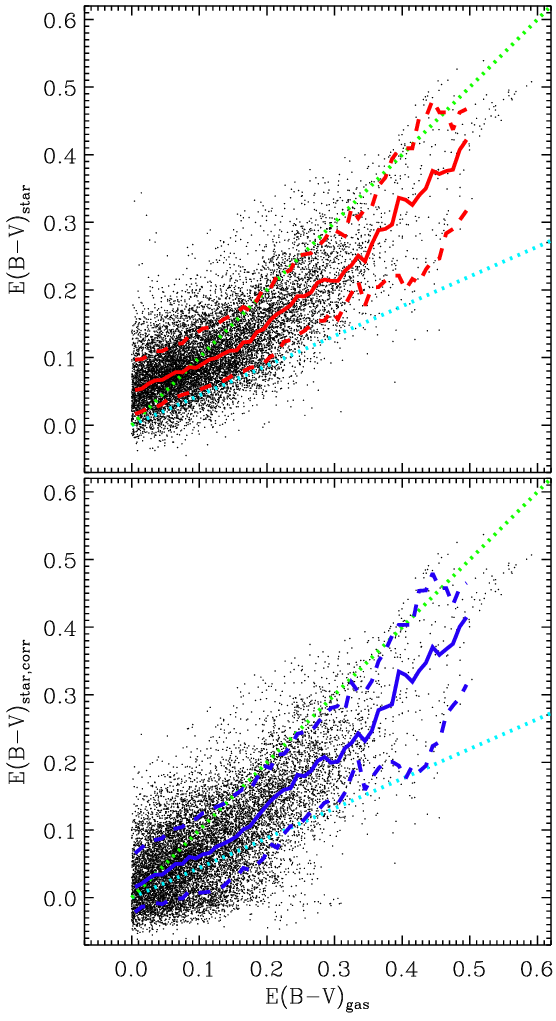} \\  
    \caption{Stellar attenuation $E(B-V)_{\rm star}$ (top) and $E(B-V)_{\rm star,corr}$ (bottom) versus gas attenuation $E(B-V)_{\rm gas}$. The red (or blue) solid and dashed lines mean the sliding medians and 16th–84th percentile distributions, respectively. 
    The green and cyan dotted lines are the one-to-one relation and the $E(B-V)_{\rm star}$ = 0.44 $E(B-V)_{\rm gas}$ relation \citep{cal00}.}
    \label{fig-ebv-comp}
\end{figure}

Based on the results of SFR indicator calibration, we can eventually determine the color excess toward stellar light as
  \begin{equation}
  E(B-V)_{\rm star} = \frac{2.5}{k_{\rm FUV}}~\log \left(\frac{L_{\rm FUV+W3,offset}}{L_{\rm FUV}}\right),
  \end{equation}
  where $k_{\rm FUV}$ is from the \citet{cal00} reddening curve.
There have been reports of variations in the reddening curves from galaxy to galaxy or from region to region \citep[e.g.,][]{wil11,sal18,tac22}. 
The impact of choosing a different curve can be considerable in the FUV band and $E(B-V)_{\rm star}$, but it is difficult to provide a detailed discussion on this impact in our study.
In the top panel of Figure \ref{fig-ebv-comp}, we find a proportional relationship between $E(B-V)_{\rm gas}$ and $E(B-V)_{\rm star}$ with a 1$\sigma$ scatter of 0.046 around the median trend. 
The narrow sequence of spaxels with $E(B-V)_{\rm gas} >$ 0.5 is driven by the fact that the majority of them (88 percent) are from one galaxy, NGC 5145.
There are some unrealistic cases (1.3 percent) with $E(B-V)_{\rm star} <$ 0.
If $L_{\rm FUV+W3,coeffi}$ is used instead of $L_{\rm FUV+W3,offset}$ when obtaining $E(B-V)_{\rm star}$, these cases disappear by definition, but the overall tendency between the gas and stellar attenuation does not change at all.
More than half of our spaxels (57 percent) lie between the line of $E(B-V)_{\rm star}$ = 0.44 $E(B-V)_{\rm gas}$ \citep{cal00} and the one-to-one line.
Remarkably, when $E(B-V)_{\rm gas}$ is small, the spaxels are heavily skewed toward $E(B-V)_{\rm star} > E(B-V)_{\rm gas}$.
The median value of $E(B-V)_{\rm star}$ at $E(B-V)_{\rm gas} \approx$ 0 is 0.051.
This at least partially originates from methodological limitations in selecting spaxels with W3 intensity $>$ 0, considering that the value of 0.051 increases as a higher S/N threshold is applied.
Another possibility is that the IR emission component attributed to old stellar populations remains substantial even by employing the conversion offset (or combination coefficient) that varies with sSFR, especially for regions with small $E(B-V)_{\rm gas}$.
The median $L_{\rm W3}$ at $E(B-V)_{\rm gas} \approx$ 0 is 10$^{5.44}$ $L_{\odot}$, $L_{\rm W3,0}$ hereafter.
Under the assumption that the W3 emission involved in pure star-forming regions can be extracted by simply subtracting $L_{\rm W3,0}$ from $L_{\rm W3}$, we calculate $L_{\rm FUV+W3,offset,corr}~(= 10^{~\textit{\textbf{b}}+2.18}~[L_{\rm FUV} + 4.63~(L_{\rm W3}-L_{\rm W3,0})])$\footnote{If $L_{\rm W3}$ is less than $L_{\rm W3,0}$, we do not apply the $L_{\rm W3,0}$ correction.}. 
We then obtain $E(B-V)_{\rm star,corr}$ based on $L_{\rm FUV+W3,offset,corr}$ and compare it with $E(B-V)_{\rm gas}$ in the bottom panel.
The median $E(B-V)_{\rm star,corr}$ at $E(B-V)_{\rm gas} \approx$ 0 is 0.013, which is much lower than 0.051.
After the $L_{\rm W3,0}$ correction, the trend line passes near the origin, but the related scatter increases from 0.046 to 0.057. 
When comparing the trend lines of the two panels, they are significantly different at $E(B-V)_{\rm gas} \lesssim$ 0.1 but almost identical at $E(B-V)_{\rm gas} \gtrsim$ 0.2.
$E(B-V)_{\rm gas}$ is used to derive $L_{\rm H\alpha,corr}$ and $E(B-V)_{\rm star}$ is related to the SFR calibration of $L_{\rm FUV+W3}$ against $L_{\rm H\alpha,corr}$. 
Therefore, $E(B-V)_{\rm gas}$ and $E(B-V)_{\rm star}$ are not completely independent quantities in this study. 
However, since the influence of $E(B-V)_{\rm gas}$ on the determination of $E(B-V)_{\rm star}$ is minimal (negligible correlations between $L_{\rm FUV+W3}$/$L_{\rm H\alpha,corr}$ and $E(B-V)_{\rm star}$), their comparison remains physically meaningful. 
Even if $E(B-V)_{\rm star}$ is calculated using previously known SFR calibration results in the literature, we still obtain similar trends.

\begin{figure}
    \centering
    \includegraphics[width=0.9\linewidth]{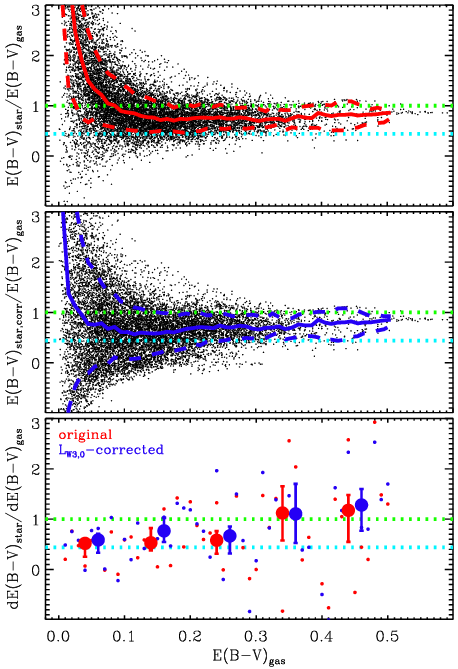} \\ 
    \caption{$E(B-V)_{\rm star}$/$E(B-V)_{\rm gas}$ (top), $E(B-V)_{\rm star,corr}$/$E(B-V)_{\rm gas}$ (middle), and the slopes of attenuation trend lines (bottom) as a function of $E(B-V)_{\rm gas}$.
    The guide values of 1.0 and 0.44 are presented as green and cyan horizontal dotted lines.
    In the top and middle panels, the solid and dashed lines indicate the sliding medians and 16th–84th percentile distributions.
    In the bottom panel, the red and blue dots are directly from the trend line before and after the $L_{\rm W3,0}$ correction, respectively.
    Each red (blue) circle is the median of ten red (blue) dots.
    The red and blue circles are properly separated for better visibility.
    Their error bars are estimated using random samplings.}
    \label{fig-ebv-slope}
\end{figure}

To more clearly demonstrate that the trend lines are curved, we plot $E(B-V)_{\rm star}$/$E(B-V)_{\rm gas}$, $E(B-V)_{\rm star,corr}$/$E(B-V)_{\rm gas}$, and the slopes of the lines as a function of $E(B-V)_{\rm gas}$ in Figure \ref{fig-ebv-slope}.
The $E(B-V)_{\rm star}$/$E(B-V)_{\rm gas}$ is positively correlated ($\rho = 0.047$) with $E(B-V)_{\rm gas}$ when limited to $E(B-V)_{\rm gas} >$ 0.15, while the $E(B-V)_{\rm star,corr}$/$E(B-V)_{\rm gas}$ is more strongly correlated ($\rho = 0.161$) even at $E(B-V)_{\rm gas} >$ 0.10.
Both attenuation ratios appear to be less meaningful in small $E(B-V)_{\rm gas}$ spaxels because of their exceptionally large scatter and the issue that the trend lines do not pass through the origin.
In the bottom panel, the red dots are from the variation of median $E(B-V)_{\rm star}$ when $E(B-V)_{\rm gas}$ increases by 0.01, which correspond almost to the instantaneous slopes of the line.
Each red circle is a representative of ten red dots, indicating the median slope in the $E(B-V)_{\rm gas}$ bin of 0.1.
The blue dots and circles are based on the trend line between $E(B-V)_{\rm star,corr}$ and $E(B-V)_{\rm gas}$.
For both trend lines, the median slopes are close to 0.44 at $E(B-V)_{\rm gas} \lesssim$ 0.3 and close to 1.0 at $E(B-V)_{\rm gas} \gtrsim$ 0.3.
This systematic change in line slope implies that the differential attenuation is influenced by the physical conditions of spaxels.

\begin{figure*}
    \centering
    \includegraphics[width=1.0\linewidth]{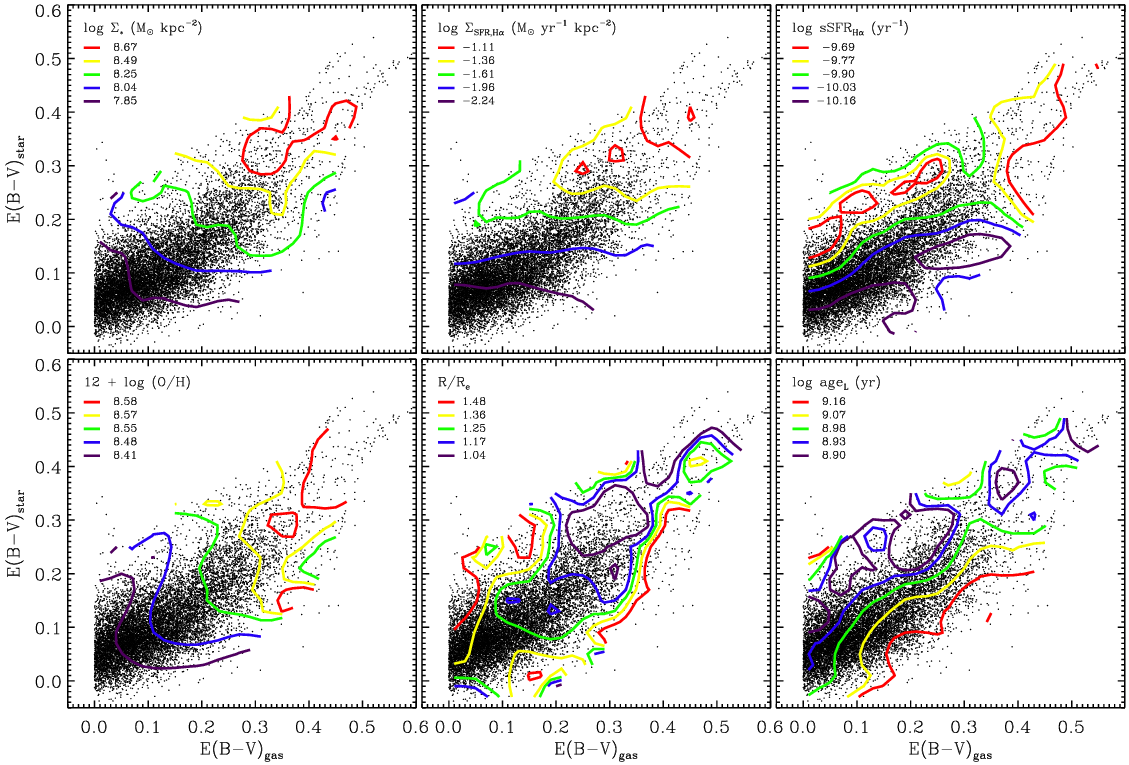} \\
    \caption{Relation between $E(B-V)_{\rm star}$ and $E(B-V)_{\rm gas}$, overlaid with contours of the several physical parameters. 
    They include $\Sigma_*$ (top-left), $\Sigma_{\rm SFR,H\alpha}$ (top-middle), sSFR$_{\rm H\alpha}$ (top-right), 12 + log (O/H) (bottom-left), $R$/$R_{\rm e}$ (bottom-middle), and age$_{\rm L}$ (bottom-right).}
    \label{fig-ebv-cont}
\end{figure*}

Figure \ref{fig-ebv-cont} illustrates how the physical parameters change in the plane of $E(B-V)_{\rm gas}$ and $E(B-V)_{\rm star}$.
The $\Sigma_*$ and 12 + log (O/H) contour lines are roughly perpendicular to the main sequence. 
The stellar mass density and gas-phase metallicity are on average higher in dustier systems (in terms of both gas and stellar attenuation), consistent with previous studies \citep[e.g.,][]{zah17,shi20,lor24}.
However, there are also reports that gas attenuation is positively correlated with stellar mass (and metallicity) but stellar attenuation exhibits a negative or negligible correlation \citep[e.g.,][]{gre20,lin20}.
Interestingly, the $\Sigma_{\rm SFR,H\alpha}$ contour lines are not vertical but horizontal, even though $E(B-V)_{\rm gas}$ is used to derive $\Sigma_{\rm SFR,H\alpha}$, increasing the correlation between them.
Considering that a similar pattern of contour lines is seen with $\Sigma_{\rm SFR,FUV+W3}$, it is evident that stellar attenuation is strongly affected by SFR surface density compared to gas attenuation.
This could be a result of the connection between high SFR density and dense/opaque ISM that reddens the stellar continuum photons (e.g., \citealt{red10,pan15}; see also \citealt{sal23}).
The sSFR$_{\rm H\alpha}$ and age$_{\rm L}$ contour lines tend to be parallel to the sequence.
This suggests that the sSFR and mean stellar age are closely linked with the differential attenuation instead of the amount of attenuation.
The high sSFR (and young age) systems are dominated by stars associated with dusty birth clouds rather than diffuse ISM and thus appear to suffer from extra stellar attenuation \citep[e.g.,][]{wil11,koy19,li21}.
We emphasize that if using $E(B-V)_{\rm star}$ from $L_{\rm FUV+W3,const}$ (i.e., without the sSFR-dependent \textit{\textbf{a}} or \textit{\textbf{b}}), the opposite trend is obtained that high sSFR ones are at the lower part of the sequence, same as fig. 4 in \citet{bel23}. 
Meanwhile, the $R$/$R_{\rm e}$ contour lines look complicated although there is a weak signal of high attenuation in central regions (see Section \ref{sec-global} for more details). 

\begin{figure*}
    \centering
    \includegraphics[width=1.0\linewidth]{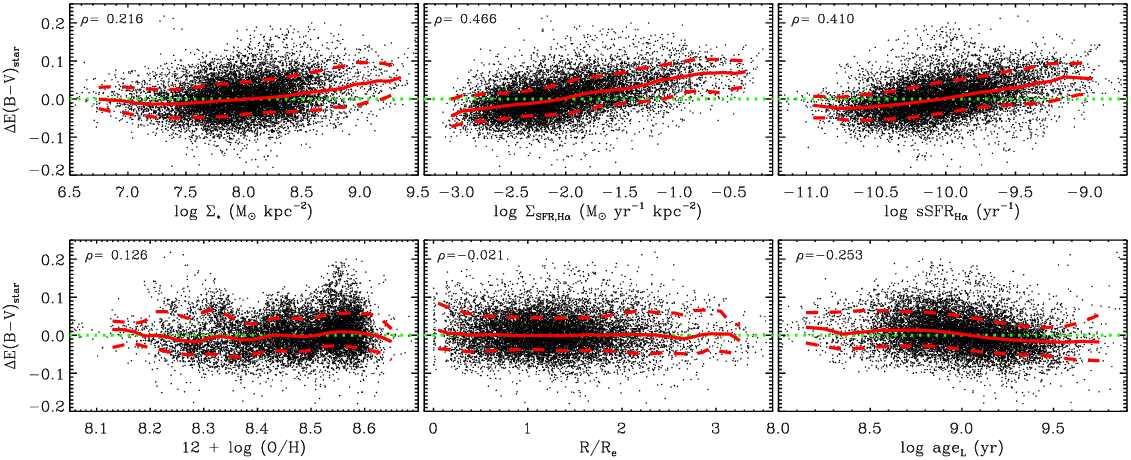} \\
    \caption{$\Delta E(B-V)_{\rm star}$ as a function of $\Sigma_*$ (top-left), $\Sigma_{\rm SFR,H\alpha}$ (top-middle) sSFR$_{\rm H\alpha}$ (top-right), 12 + log (O/H) (bottom-left), $R$/$R_{\rm e}$ (bottom-middle), and age$_{\rm L}$ (bottom-right).
    In each panel, the red solid and dashed lines stand for the sliding medians and 16th--84th percentile distributions, respectively. 
    The Spearman's $\rho$ is denoted in the top-left corner.}
    \label{fig-ebv-resi}
\end{figure*}

In order to parameterize the differential attenuation, we derive $\Delta E(B-V)_{\rm star}$ as $E(B-V)_{\rm star}$ deviation from the median trend between $E(B-V)_{\rm star}$ and $E(B-V)_{\rm gas}$, indicating how far the spaxel lies above or below the trend line, irrespective of the shape or offset of the trend line.
When $\Delta E(B-V)_{\rm star}$ is positive and negative, the stellar-to-gas attenuation ratio is generally close to 1.0 and 0.44, respectively.
Compared to a simple difference or ratio between stellar and gas attenuation, this parameter has advantages in describing the relative attenuation, especially for less dusty systems. 
Figure \ref{fig-ebv-resi} presents $\Delta E(B-V)_{\rm star}$ as a function of several physical parameters.
The $\Delta E(B-V)_{\rm star}$ is strongly correlated with $\Sigma_{\rm SFR,H\alpha}$ and sSFR$_{\rm H\alpha}$ ($\rho =$ 0.466 and 0.410).
The modest correlations with $\Sigma_*$ and age$_{\rm L}$ ($\rho =$ 0.216 and $-$0.253) seem to be a secondary outcome that arises from the $\Sigma_{\rm SFR,H\alpha}$-$\Sigma_*$ and sSFR$_{\rm H\alpha}$-age$_{\rm L}$ relationships, respectively.
The correlation with $\Sigma_*$ (age$_{\rm L}$) becomes much weaker when fixing $\Sigma_{\rm SFR,H\alpha}$ (sSFR$_{\rm H\alpha}$).
In contrast, the correlation with $\Sigma_{\rm SFR,H\alpha}$ (sSFR$_{\rm H\alpha}$) remains strong regardless of fixing $\Sigma_*$ (age$_{\rm L}$).
Between $\Sigma_{\rm SFR,H\alpha}$ and sSFR$_{\rm H\alpha}$, the correlation with $\Sigma_{\rm SFR,H\alpha}$ at fixed sSFR$_{\rm H\alpha}$ is stronger than the correlation with sSFR$_{\rm H\alpha}$ at fixed $\Sigma_{\rm SFR,H\alpha}$.
We therefore conclude that differential attenuation is primarily regulated by SFR density and is followed by sSFR.
\citet{qin19} argued that attenuation ratio is controlled by sSFR density, but the correlation with $\Sigma_{\rm sSFR,H\alpha}$ ($\rho =$ 0.343) is milder than that with $\Sigma_{\rm SFR,H\alpha}$ or sSFR$_{\rm H\alpha}$ in this study. 
Note that these correlations diminish substantially if we simply use $E(B-V)_{\rm star}-E(B-V)_{\rm gas}$ or $E(B-V)_{\rm star}$/$E(B-V)_{\rm gas}$.
When using $E(B-V)_{\rm star}$/$E(B-V)_{\rm gas}$ for the subset with $E(B-V)_{\rm gas} \gtrsim$ 0.15, we can obtain similar results although statistically less significant.
\citet{wil11}, \citet{koy19}, and \citet{li21} also reported that high sSFR systems show extra stellar attenuation.
However, the results of \citet{koy19} and \citet{li21} differ from ours in detail because they found that attenuation ratios are well correlated with mass rather than with SFR.
This implies that the relationships between attenuation and other physical properties are complex, and it is important to validate the results repeatedly using diverse samples and methodologies.

\subsection{Local and Global Effects}\label{sec-global}

\begin{figure*}
    \centering
    \includegraphics[width=0.875\linewidth]{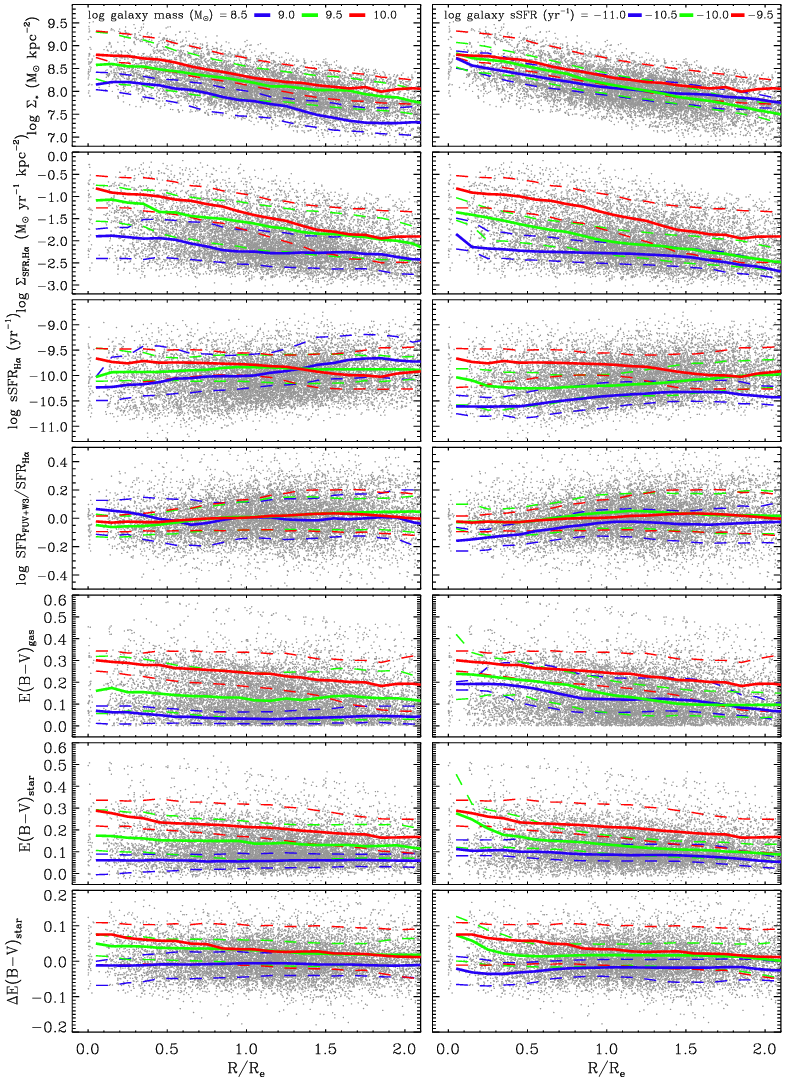} \\
    \caption{From top to bottom, $\Sigma_*$, $\Sigma_{\rm SFR,H\alpha}$, sSFR$_{\rm H\alpha}$, SFR$_{\rm FUV+W3}$/SFR$_{\rm H\alpha}$, $E(B-V)_{\rm gas}$, $E(B-V)_{\rm star}$, and $\Delta E(B-V)_{\rm star}$ as a function of $R$/$R_{\rm e}$.
    Different galaxy radial profiles are overplotted in the left and right panels, color-codded by their mass and sSFR ranges, respectively.
    The solid and dashed lines indicate the median profile and its dispersion.}
    \label{fig-profile}
\end{figure*}

\begin{figure*}
    \centering
    \includegraphics[width=0.9\linewidth]{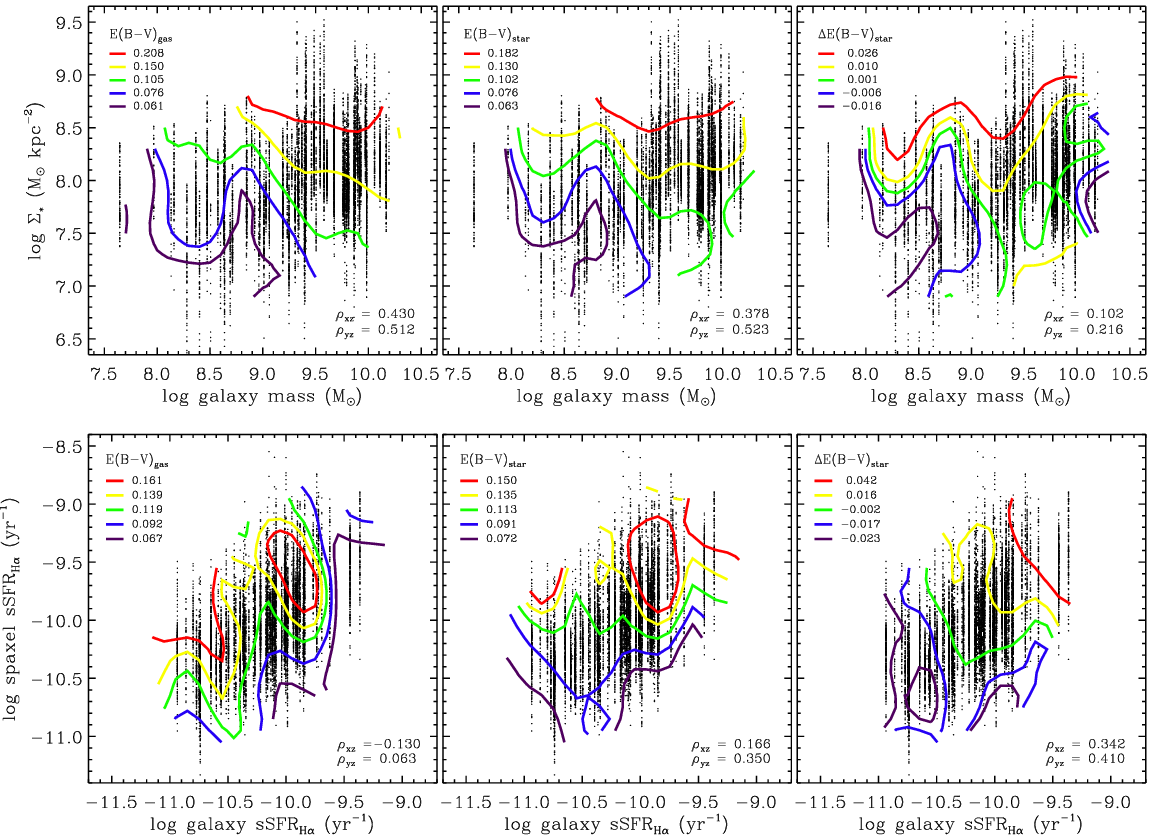} 
    \caption{Top: galaxy mass and $\Sigma_*$ diagrams of spaxels. $E(B-V)_{\rm gas}$ (left), $E(B-V)_{\rm star}$ (middle), and $\Delta E(B-V)_{\rm star}$ (right) contour lines are overlaid.
    In each panel, the $\rho_{xz}$ ($\rho_{yz}$) value denote the correlation coefficient between the attenuation and x-axis (y-axis) parameters.  
    Bottom: Same as the top panels, but for galaxy sSFR and spaxel sSFR diagrams.}
    \label{fig-global}
\end{figure*}

Figure \ref{fig-profile} exhibits the radial distribution of key parameters for our spaxels.
The median radial profiles are overlaid using the spaxels that inhabit galaxies within a specific range of mass and sSFR.
In the left panels, to investigate the galaxy mass dependence along the SFMS, the profiles are based on the galaxies with mass ranging from 10$^{8.5}$ to 10$^{10.0}$ $M_{\sun}$ (three bins at intervals of 0.5 dex) at fixed sSFR of 10$^{-10.0}$--10$^{-9.5}$ yr$^{-1}$. 
In the right panels, to check the sSFR dependence across the SFMS, they are from the galaxies with sSFR from 10$^{-11.0}$ to 10$^{-9.5}$ yr$^{-1}$ at fixed mass of 10$^{9.5}$--10$^{10.0}$ $M_{\sun}$.  
We define the bin boundaries to uniformly maximize the number of galaxies in each bin, resulting in 7--8 galaxies per bin.
Note that the radial profiles could be uncertain at $R$/$R_{\rm e}$ $\lesssim$ 0.5 or $\gtrsim$ 1.5 because of the beam smearing effects (typical $R_{\rm e}$ of our galaxies = 12--22$\arcsec$ and spatial resolution of our datasets = 7$\arcsec$.5) and the low completeness of spaxels (more than half of the entire spaxels not satisfying our selection criteria), respectively.

The $\Sigma_*$ and $\Sigma_{\rm SFR,H\alpha}$ profiles are elevated overall in relatively massive galaxies.
Their slopes change systematically so that the sSFR$_{\rm H\alpha}$ profile has a negative (positive) radial gradient in the high (low) mass bin.
Similar results are obtained through the sSFR-binned profiles.
These sSFR$_{\rm H\alpha}$ profile variations are compatible with previous studies (e.g., \citealt{ell18,mun24}; but see also \citealt{lin24}).
In the SFR$_{\rm FUV+W3}$/SFR$_{\rm H\alpha}$ profile (SFR$_{\rm FUV+W3}$ from $L_{\rm FUV+W3,offset}$), the galaxy mass and sSFR effects are marginal in the sense that at 0.5--1.5 $R$/$R_{\rm e}$, the 1$\sigma$ scatter of spaxels is 0.135 dex, but the median difference between the maximum and minimum values is only 0.027 (0.073) dex in the mass (sSFR) bins. 
It supports that our SFR calibration recipe of $L_{\rm FUV+W3,offset}$ performs well for a wide range of dusty star-forming galaxies.  

On the other hand, we recognize that the $E(B-V)_{\rm gas}$ and $E(B-V)_{\rm star}$ profiles are also boosted in massive galaxies.
Both gas and stellar attenuation profiles tend to be centrally concentrated in the high mass bin, while they are nearly flat in the low mass bin.
\citet{nel16} and \citet{gre20} found such a mass dependence of gas attenuation profile as well.
However, \citet{gre20} reported that stellar attenuation profiles depend little on galaxy mass, different from our result.
This inconsistency is possibly due to the different data used to derive stellar attenuation: FUV and MIR intensities in our case versus optical spectra in theirs. 
The galaxy mass and sSFR dependence of $\Delta E(B-V)_{\rm star}$ is substantial: at 0.5--1.5 $R$/$R_{\rm e}$, 1$\sigma$ scatter of spaxels = 0.045 dex and median difference between the maximum and minimum values in the mass (sSFR) bins = 0.040 (0.052) dex.
The radial gradient of $\Delta E(B-V)_{\rm star}$ also clearly varies: it is negative (almost zero) in the high (low) mass and sSFR bins.
We thus expect that the differential attenuation of spaxels is affected by the global properties of their host galaxy.

In Figure \ref{fig-global}, to distinguish between the local and global effects on $E(B-V)_{\rm gas}$, $E(B-V)_{\rm star}$, and $\Delta E(B-V)_{\rm star}$, we draw their contour lines in the planes of $\Sigma_*$ versus galaxy mass and spaxel sSFR$_{\rm H\alpha}$ versus galaxy sSFR$_{\rm H\alpha}$, and calculate the $\rho$ values between attenuation and local/global parameters.
The contour patterns appear to be complicated, but the attenuation parameters have a stronger correlation with $\Sigma_*$ than with galaxy mass.
The $E(B-V)_{\rm star}$ and $\Delta E(B-V)_{\rm star}$ are also highly correlated with spaxel sSFR rather than with galaxy sSFR.
The weak negative correlation between $E(B-V)_{\rm gas}$ and galaxy sSFR is mainly due to a few galaxies with high sSFR ($\gtrsim$ 10$^{-9.5}$ yr$^{-1}$) but low gas attenuation.
Considering their median 12 + log (O/H) $<$ 8.4, they are likely galaxies where intense star formation is fueled by metal-poor gas \citep[e.g.,][]{ly15,gao18}.
Although the correlations are overall stronger with the local parameters than with the global parameters, it is difficult to ensure that the local effect is more dominant in the sense that we cannot rule out that such a difference might be just because the attenuation parameters are also local values.
A trend with the global parameter is observed even when fixing the local parameter and vice versa, particularly in the top-left, top-middle, and bottom-right panels.
It suggests that the attenuation parameters are significantly influenced by both local and global effects.
Regardless of whether it is a local or global property, mass seems to be more important in determining the amount of attenuation than sSFR, while sSFR plays a more crucial role in determining the differential attenuation than mass.

\section{SUMMARY}\label{sec-sum}

We select 74 nearby star-forming galaxies having \textit{GALEX} FUV, CALIFA optical, and \textit{WISE} 12 \micron{} (W3) data and reconstruct the optical IFS datacubes to be spatially matched with the FUV and MIR images.
For $\sim$12,000 spaxels, we calibrate the hybrid SFR indicator of FUV and W3 combined luminosity to the dust-corrected H$\alpha$ luminosity and obtain the stellar attenuation.
We compare the gas and stellar attenuation and inspect the differential attenuation as a function of spaxel and galaxy properties.
Our main results are as follows.
\begin{enumerate}
    \item The simple linear combination of FUV and W3 luminosities follows a one-to-one relationship with the dust-corrected H$\alpha$ luminosity, although their relation is slightly tilted.
    \item Using the combination coefficient or conversion offset that varies with sSFR, the relations become tighter and have minimal dependence on physical parameters, providing more robust SFR recipes.
    \item In the plane of gas versus stellar attenuation, the median trend line is curved, with its slope ($\approx$ stellar-to-gas attenuation ratio) changing from 0.44 to 1.0 as dust attenuation increases.
    \item The differential attenuation, quantified by how much stellar attenuation deviates from the median trend line, shows the strongest correlation with the SFR density and sSFR, which can be well explained by the variation in relative importance of birth clouds and the diffuse ISM.
    \item The differential attenuation is influenced by galaxy parameters even when fixing spaxel parameters, suggesting that there are both local and global effects on the differential attenuation.
\end{enumerate}
Our study has substantially enlarged the sample size of galaxies compared to previous studies, but is still restricted to galaxies near the star-forming main sequence. 
For a more comprehensive study encompassing a variety of galaxies, high spatial resolution, deep survey observations need to be conducted, particularly in the UV and IR bands.

\section*{Acknowledgements}
We thank the anonymous referee for comments that helped to improve the paper. 

This research was supported by the Korea Astronomy and Space Science Institute under the R\&D program (project No. 2025-1-831-01) supervised by the Korea AeroSpace Administration.
This research was partially supported by the Australian Research Council Centre of Excellence for All Sky Astrophysics in 3 Dimensions (ASTRO 3D), through project No. CE170100013.
J.H.L. and H.J. acknowledge support from the National Research Foundation of Korea (NRF) grant funded by the Korea government (Ministry of Science and ICT, MSIT) (No. 2022R1A2C1004025).
M.P. acknowledges support from the NRF grant funded by the MSIT (No. 2022R1A2C1004025 and 2022R1C1C2006540).
S.O. acknowledges support from the NRF grant funded by the MSIT (RS-2023-00214057).

This study uses data provided by the Calar Alto Legacy Integral Field Area (CALIFA) survey (\href{http://califa.caha.es/}{http://califa.caha.es/}).
Based on observations collected at the Centro Astronómico Hispano Alemán (CAHA) at Calar Alto, operated jointly by the Max-Planck-Institut für Astronomie and the Instituto de Astrofísica de Andalucía (CSIC).
\facilities{\textit{WISE}, \textit{GALEX}}
\software{\ppxf{} \cite[]{cap23}}

\end{document}